# Origin of mixed anisotropy in crystalline Permalloy and amorphous Cobalt thin films individually deposited on Si substrate

Kirti Kirti, Baisali Ghadai, Abinash Mishra, Rahulkrishnan R, Sucheta Mondal*

*Shiv Nadar Institution of Eminence, Greater Noida, Delhi NCR, UP-201314*

*sucheta.mondal@snu.edu.in

## Abstract

Magnetic anisotropy (MA) plays a crucial role in deciding both static and dynamic behaviour of magnetic thin films. It controls various phenomena, such as magnetization reversal, domain formation, domain-wall motion, spin-wave generation, and spin-wave propagation etc. We investigate the mixed anisotropies in face-centred-cubic Permalloy (fcc-Py) and amorphous Cobalt (a-Co) thin films deposited *via* rf magnetron sputtering on Si (100) substrate with thicknesses, $d$ = 5-125 nm and $t$ = 5-150 nm, respectively. X-ray diffraction technique, atomic force microscopy, and vibrating sample magnetometry are employed to study the structural, morphological, and magnetic properties. We adopt a qualitative approach to understand the nature of different anisotropies present in both materials. Mixed anisotropies evolve with film thicknesses for both fcc-Py and a-Co films. The role of growth conditions in the emergence of specific anisotropies is discussed in detail. An alteration of the magnetization easy axis from the conventional in-plane orientation is evidenced due to the collective influence of these mixed anisotropies. Based on the dominance of anisotropy components, their origin, and the direction of magnetization tilt, we categorize our samples as belonging to specific regimes. Introduction of magnetization tilt has been proven to be an extremely innovative way to improve the performance of spintronic devices so far. The one-to-one comparison between a sputter-deposited crystalline and an amorphous magnetic material could be beneficial for building a stronger foundation for that.

# 1 Introduction

Magnetic anisotropy (MA) is a crucial factor controlling the shape of the hysteresis loops for strongly magnetic substances. The energy associated with MA directs the magnetization along a certain orientation called the easy axis. Presence of more than one easy axis is typical.[1] Magneto-crystalline anisotropy (MCA) is an example of intrinsic magnetic anisotropy that originates from spin-orbit interaction in crystalline materials. However, shape anisotropy (SA), magnetostrictive anisotropy (MsA) etc. are the external sources of anisotropies. There are several other factors which may introduce changes (permanent or transient) in the anisotropy of any magnetic substance, *i.e.*, surface roughness,[2] hybridization at the interfaces,[3] inter-layer and intra-layer couplings,[4][5][6] elemental compositions,[7] artificial patterning,[8] irradiation with laser and ions,[9][10] anisotropies due to annealing,[11] application of a magnetic field during sample growth,[12] application of external strain on the substrate,[13] and electric field in the material,[14] etc. It is customary to address this category as 'induced' these days. For a nanoscale magnetic system, more than one of these anisotropies is usually present, though any one may become predominant in a special circumstance. MCA determines the direction of spontaneous magnetization in the demagnetized state in the absence of other competitors. The magnitude and direction of MCA are material specific. For infinite thin films, the in-plane (IP) SA apparently forces the magnetization to lie in the plane of the film. This depends on the shape factor of the film ($N_{zz} = 1$) and the saturation magnetization ($M_s$) of the material. Additionally, growth-induced (GI) anisotropies can be observed in thin films. For example, a film deposited with a certain oblique angle between the target and the substrate inside the deposition chamber may accommodate GI shape anisotropy (GI-SA) due to tilted nanostructures (like nanopillars, nano-columns or nano-sheets etc.). The strength and orientation of such anisotropy depend on deposition geometry, chamber conditions, film thickness, and magnetization of the materials, etc. This prefers to orient the magnetization along the long axis of the columns (or nanostructures) formed due to the adhesion of atoms in specific arrangements. If the material is magnetostrictive in nature, an applied mechanical stress alters the domain structure and creates MsA. In the absence of additional applied stress, this GI stress could be generated from the high growth rate, a change in the Argon (Ar) pressure, thermal annealing, and lattice mismatch between the magnetic layer and nonmagnetic underlayer etc.[15][16][17][18]

A brief overview of the role of various anisotropies in controlling and improving the functionality of the magnetic and spintronic devices will be helpful for the readers.

Anisotropy can influence magnetic phenomena on various length and time scales. The slowest phenomenon in the quasistatic- and dynamic categories is the formation and motion of the domain wall. Domain wall energy is mostly influenced by the exchange energy and the anisotropy energy. Moreover, the competition between the two decides the type of wall. The effective wall thickness ($\delta$) depends directly on the effective anisotropy constant ($K_{\text{eff}}$), *i.e.* $\delta \propto \sqrt{\frac{A}{K_{eff}}}$.[19] Here, $A$ is the exchange constant. The velocity of the domain wall ($v$) and the anisotropy are also correlated in the following manner: $v \propto 1/\sqrt{K_{eff}}$.[20] MA also decides whether other magnetic textures and topological solitons, such as vortex, hopfions, skyrmion, skyrmionium, etc.[21][22][23][24] can be stabilized or not. By tuning the magnitude and nature of MA, several parameters associated with gyration dynamics (*i.e.,* amplitude, frequency, and mode formation, etc.) of these solitons can be controlled. A plethora of studies have been reported in which researchers have shown a direct correlation between MA and magnonic band structure for thin films and artificial magnonic crystals.[25] In the nanosecond regime, frequency[26] and lifetime[27][28] of magnons (*i.e.*, quanta of spin waves) can be modulated with the effective anisotropy in the system. A recent report suggests that thoughtful engineering of the anisotropy in nanoscale magnetic material facilitates spin-wave transmission *via* a programmable domain wall.[29] Most of the above-mentioned devices are still in prototype form, whereas magnetic switching devices have been commercialized, and their operation is also governed by MA. Storing and processing of magnetic information rely on efficient switching of magnetization states, which can be altered by external stimulations, such as magnetic field pulse, spin torque, ultrafast optical pulses etc. The length scale may vary according to the requirement. However, most of them are activated in less than a nanosecond time scale. These stimuli modify the magnetic state of the free layer that is stabilized by the effective anisotropy in the device before and after the switching operation. The thermal stability factor in nanomagnetic devices is directly proportional to $M_s$ and MA. Material with higher anisotropy is preferred for designing thermally robust devices. However, the write current increases, hence low damping is necessary to reduce the excess power consumption in such a case. The most common stack design for switching devices is based on the perpendicularly magnetized free layer and fixed layer.[30] There are also reports available on spin-torque-induced switching of IP magnetized nanomagnets.[31] However, the requirement of extra energy to cross the barrier height between the opposite spin states creates a bottleneck for both systems. A few years back, L. You *et al.* proposed wedge-shaped devices in which tilted anisotropy can facilitate deterministic switching without applying the symmetry-breaking field that does not have microelectronic compatibility.[32] Apart from this, a celebrated mechanism is the laser-induced modification of MA. It is the backbone of heat-assisted magnetic recording technology and all-optical switching phenomena.[9][33] The influence of anisotropy on the ultrafast time scale has also been studied, where the researchers demonstrated that the magnetization along the hard axis relaxes faster than the easy axis case due to large electron-phonon spin-flip scattering for magnetization.[34]

Despite tremendous success in this front, we identify a gap in the existing literature that needs to be addressed with experimental evidence, *i.e.*, mapping of inherent anisotropy components present in sputter-deposited crystalline and amorphous magnetic films with thicknesses lying between the ultrathin and thick regimes. Permalloy being a softer magnet and Cobalt being relatively harder magnet are technological important material for the spintronic devices, and both the material serve important and individual role in the devices' performance. In order to fabricate energy efficient devices a thorough discussion on the underlying mechanisms on the fundamental point of view with experimental evidence is necessary. However, to best of our knowledge this much rigorous discussion on each anisotropic component in two completely different magnetic materials in terms of order and magnetic properties is not reported. Additionally, the emergence and evolution of tilted magnetic anisotropy with thickness of these films have not been studied, in a seed-free environment. We thoroughly characterize face-centred-cubic Permalloy (fcc-Py) and amorphous Cobalt (a-Co) thin films deposited *via* rf magnetron sputtering on Si (100) substrate in an intermediate thickness range. A vibrating sample magnetometry (VSM) is exploited to obtain IP and out-of-plane (OOP) hysteresis loops for the films to study the mutation of coercivity ($H_c$), remanence ($M_r$), saturation field ($H_s$), $M_s$, and effective anisotropy ($K_{eff}$). We deconvolute the anisotropy contributions in these films with a qualitative approach, and their origin is discussed in detail.

## 2 Experimental Details

Py ($Ni_{80}Fe_{20}$) and Co thin films with thicknesses $d$ = 5-125 nm and $t$ = 5-150 nm are deposited respectively on Silicon substrates (Si with (100) orientation) using rf magnetron sputtering at room temperature. The deposition condition is as follows: base pressure better than $7\times10^{-7}$ Torr, source power of 80 W, working pressure of 7 mTorr, and Ar flow rate of 40 SCCM. All the thin films are deposited with an oblique angle of about 45° with respect to the substrate normal, as denoted by $\alpha$ in Figure 1a, while the substrate is rotated at nearly 2 rpm (fixed for the current chamber configuration). This slow rotation reduces the shadowing effect without completely removing it. The explanation for this is discussed later. The uncapped films (see Figure 1b-c) are used for structural-, electrical-, and magnetic characterizations. The contribution from the top oxide layer is negligible in altering the sample properties demonstrated in this paper. The samples with thicknesses more than 50 nm are tested using the profilometry [see supplementary information, Figure 1S]. The deposition rate (about 0.8 Å/sec) and thickness are estimated accordingly for all other samples. To investigate the crystal structure of these thin films, grazing-angle incidence X-ray diffraction (GI-XRD) is exploited. The surface topology of these samples is studied by the atomic force microscopy (AFM) technique in the tapping mode. The static magnetic properties of these samples are recorded in VSM measurements at room temperature. It is important to mention that both materials have the Curie temperature higher than the room temperature.

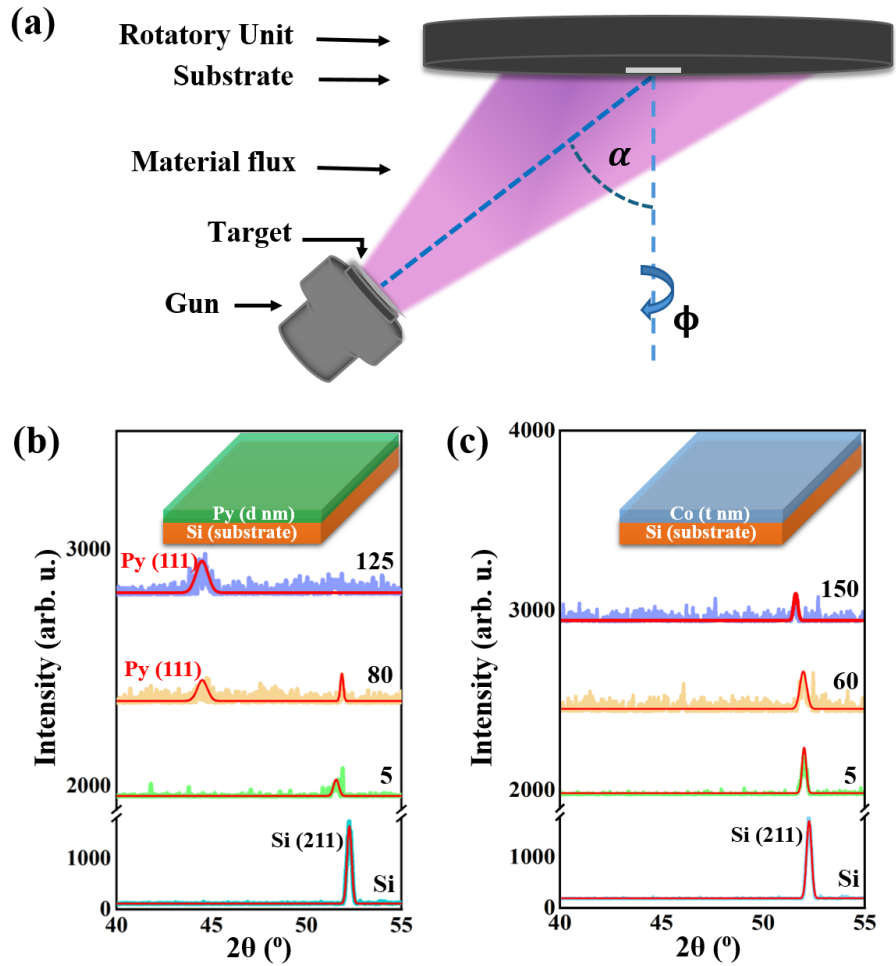

FIGURE 1: (a) Schematic diagram of the deposition geometry, $\alpha$ and $\phi$ are the angle of incident flux with respect to the normal of the substrate plane and angle of in-plane substrate rotation, respectively. The XRD spectra for (b) fcc-Py ($d$ = 5, 80, and 125 nm) and (c) a-Co ($t$ = 5, 60, and 150 nm) films, along with the Si substrate obtained in GI mode. The respective sample stacks are presented inside the figures. The numbers represent the thickness in nm. The (hkl) values are mentioned near the individual peaks. Solid red curves correspond to the fitted peaks using a Gaussian function.

## 3 Results and Discussion

### 3.1 Structural Characterization

The background-subtracted XRD spectra for the samples are presented in Figure 1b-c. The peak intensities with respect to the baseline are retained as obtained during the XRD measurements in these plots. For the Si substrate, we have obtained a peak near 52.00° that corresponds to the (211) plane. This is a forbidden peak, and it may arise due to multiple diffractions in GI mode.[35] This peak is present for most of the magnetic samples. The bulk XRD data confirms the presence of the (100) plane from the peak positioned at 69.30° [see supplementary information, Figure 2Sc]. No other peaks are found in Py samples with $d < 80$ nm, which corroborates some of the earlier reports. Films below this thickness range might not show any peak in XRD due to tiny, randomly oriented crystallites (poor crystallinity).[36] The development of a prominent peak at 44.50°, for films with $d = 80$ and 125 nm, confirms the fcc phase of Py with (111)[16] plane. To analyze the XRD spectra further, we have fitted it with a Gaussian peak function. The FWHM extracted from the fitting ranges between 0.57° and 0.67°. After applying the Debye-Scherrer formula [see supplementary information, Figure 2S], the crystallite size ($D$) is calculated for the individual peak ranges between 12.7 and 14.8 nm. Table 1 contains information about these parameters. We anticipate that for the low-thickness samples, this size is even smaller. Though the films could be weakly crystalline, the lack of enough material creates hurdles in obtaining a sufficient diffraction signal. Additionally, randomization of the fragmented crystallites reduces the average signal strength, due to which the films behave like XRD-amorphous material. The lattice constants for Py (0.35 nm for the fcc phase) and Si (0.54 nm for the cubic phase) are calculated by using Bragg's law [see supplementary information]. The huge lattice mismatch, $\eta = -54\%$, supports the growth of fcc-Py with different orientations on simple cubic-Si substrate. If there is no lattice mismatch between the film and the substrate, film should also grow with the similar orientation as the substrate. Since, there is substantial lattice mismatch which justifies the presence fcc (111) orientation of Py over the Si (100) substrate. The absence of any peak apart from the one around 52.00° in the XRD spectra for Co films indicates the amorphous nature (see Figure 1c).

TABLE 1: Peak positions, FWHM, peak intensity, and calculated crystallite size from the XRD spectra obtained for Py ($d = 80$ and 125 nm) and Si substrate are listed.

| Sample | $2\theta$ (°) | FWHM (°) | Intensity (arb. u.) | Crystallite size, $D$ (nm) |
|---|---|---|---|---|
| Py ($d = 80$ nm) | 44.50 | 0.57 | 12.46 | 14.8 |
| Py ($d = 125$ nm) | 44.50 | 0.67 | 92.50 | 12.7 |
| Si substrate | 69.30 | 0.19 | 230181.10 | 48.2 |

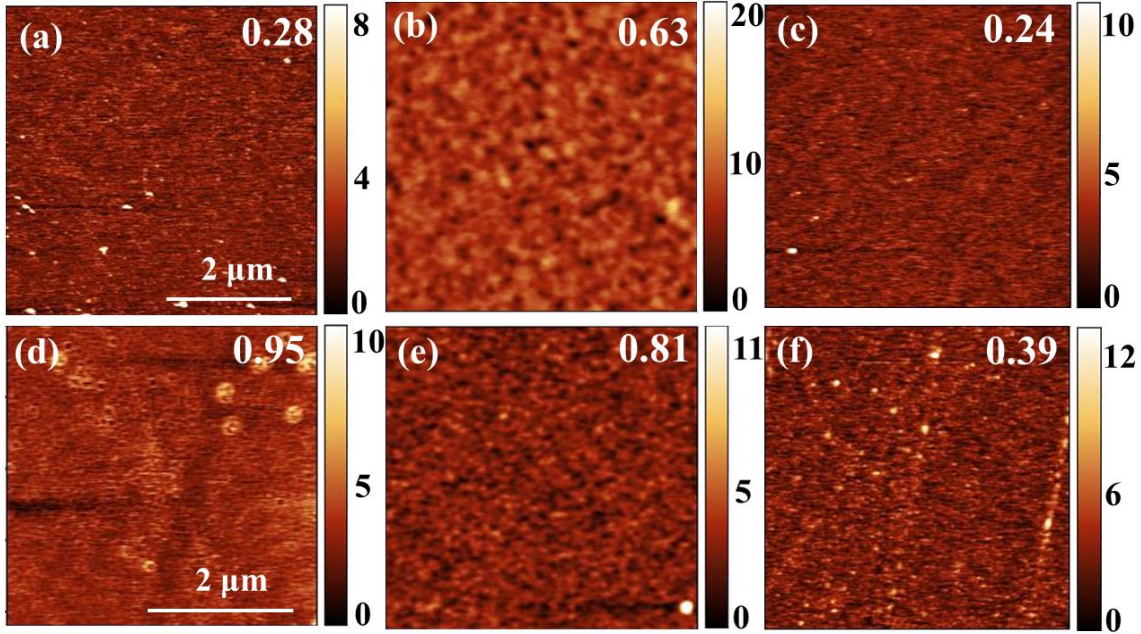

FIGURE 2: AFM images of (a-c) fcc-Py ($d$ = 5, 50, and 90 nm), and (d-f) a-Co ($t$ = 5, 40, and 90 nm). The scan area presented in the figures is $5 \mu m \times 5 \mu m$ for all the samples except one Co film ($t$ = 5 nm). Length scales are provided for reference. The colour bars indicate the height profile in nm. The numbers inside the figures correspond to the average roughness of the sample in nm.

Atomic force microscopy images reveal the film surfaces to be uniform, and Figure 3S represents a reference image for the bare Si substrate as well. The average roughness ranges between 0.24 nm and 0.63 nm for the measured Py films (Figure 2a-c). For Co samples, this value ranges between 0.39 nm and 0.95 nm, which is slightly higher than Py (Figure 2d-f). The minimum roughness value of Co is 62.5% higher than that of Py. Similarly, the maximum roughness for Co is 50.7% higher than that of Py. This higher roughness of Co can be associated with the higher surface energy of Co[37] and hence leading to the greater tendency of island and/or cluster formation at the early growth stages. The roughness variation for most of the films values are independent of thickness. However, in case of Co it can be seen that roughness is slightly decreasing with increasing thickness. At the initial growth stages there can be island and/or cluster formation which might lead to higher roughness values. As the film thickens, surface diffusion can lead to smoother films. Though the films are uncapped, we have not observed significant island formation due to oxidation in most of the films.

### 3.2 Investigation of Magnetic Properties

Figure 3a shows the normalized IP and OOP hysteresis loops for fcc-Py with varying $d$. For $d$ = 5-25 nm, the IP and OOP loops mostly coincide with a small coercive field about 25 Oe. The $H_s$ (IP) increases from 6700 Oe to 7350 Oe, which indicates that the easy axis does not confine within the sample plane. This observation also does not corroborate with conventional hysteresis behaviour for Py films (or other magnetic systems) with similar thicknesses.[38][39] To magnetize these samples in the IP and OOP directions, the same amount of energy is required to be provided by the external magnetic field during the measurement.

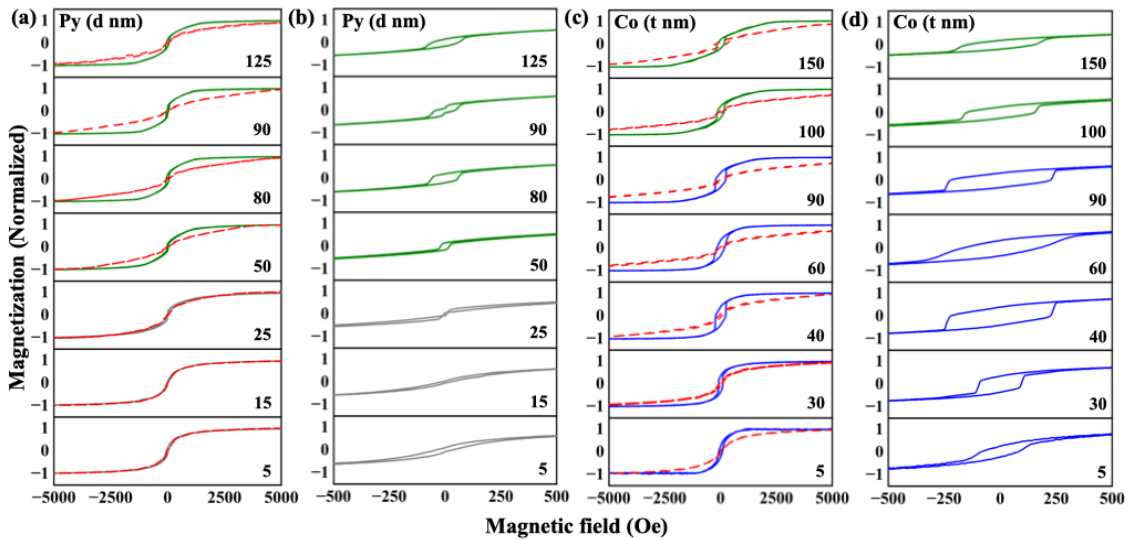

FIGURE 3: (a-d) Normalized hysteresis loops for fcc-Py and a-Co films. IP (solid line) and OOP (dashed line) loops for fcc-Py and a-Co films are presented in (a) and (c), respectively, in the range of ±5000 Oe. (b) and (d) represent the normalized IP loops in the range of ±500 Oe for both materials. Different regimes are presented in different colours: for (a-b) RT (5 nm ≤ $d$ ≤ 25 nm) in grey and ST (50 nm ≤ $d$ ≤ 125 nm) in olive; for (c-d) MIP (5 nm ≤ $t$ ≤ 90 nm) in blue and ST (100 nm ≤ $t$ ≤ 150 nm) in olive. The numbers inside the plots represent the thickness in nm.

This suggests that the favourable direction is an intermediate orientation. This is designated as the 'robustly tilted' (RT) regime here onwards. For the samples with $d > 25$ nm, OOP curves appear to be more tilted. For $d = 50$ nm, $H_s$ (IP) = 7300 Oe and $H_s$ (OOP) = 8300 Oe. This difference increases with increasing film thickness up to $d = 90$ nm. A so-called 'transcritical loop' is evidenced for the films with $d \geq 50$ nm.[15] The thickness range, 50 nm ≤ $d$ ≤ 125 nm, is termed as 'subtly tilted' (ST) regime. We observe a slight deviation from the trend for $d = 125$ nm: $H_s$ (IP) increases. Though the magnetization primarily has IP alignment however there is a presence of a weak OOP component. The core of the IP hysteresis loop is presented in Figure 3b for fcc-Py films. The shape is inverted for the first regime, which is most prominent for $d = 25$ nm. A 'pot belly'-like feature is visible for the loops between $d = 50$ nm and 80 nm. Interestingly, for $d = 90$ nm, the core of the IP loop appears to be 'wasp-waist' like. This is a composite magnetization curve-like feature where loops are formed due to the presence of two materials with conventional hysteresis behaviour and dissimilar coercivity values. Quantitative analysis of these possible effects is beyond the scope of this work. The Figure 3c-d represents the hysteresis loops for a-Co films. The samples are more coercive than fcc-Py films with $H_c$ about 150 Oe. This is conventional for Co-like harder magnets. More detailed discussion on the quantitative analysis can be found in the supplementary information [see Figures 6S and 7S]. There is no evidence of RT regime in the low thickness range. From

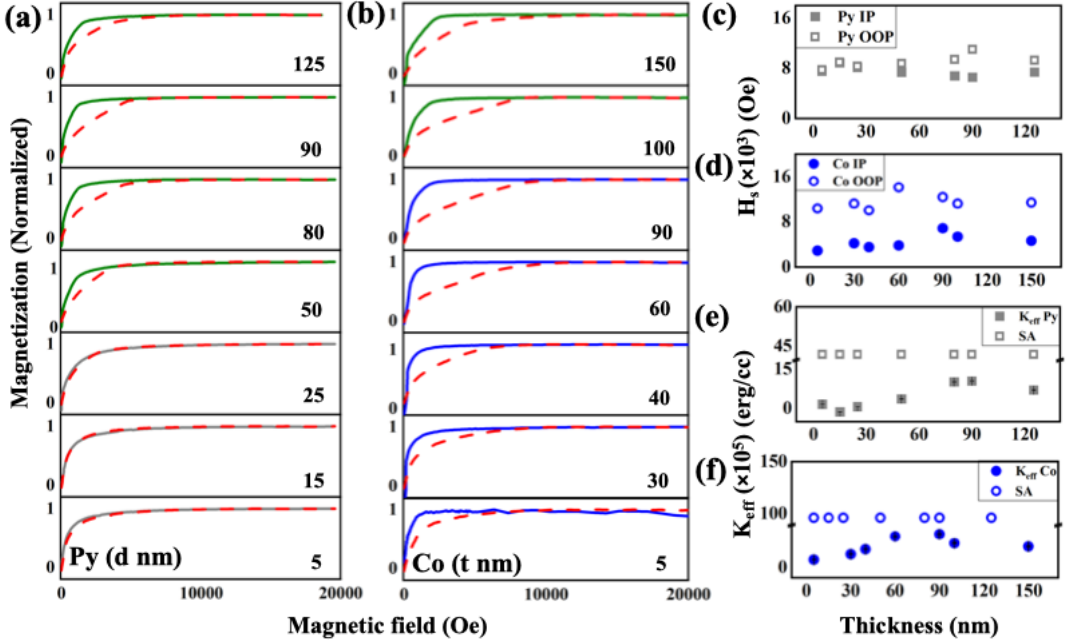

FIGURE 4: Normalized IP (solid line) and OOP (dashed line) initial curves for (a) fcc-Py and (b) a-Co films, respectively. Variation of $H_s$ with the thickness of the (c) fcc-Py and (d) a-Co, respectively, obtained from initial curves. Variation of $K_{eff}$ with the thickness of the (e) fcc-Py and (f) a-Co, respectively. $K_{eff}$ values obtained from the area method and $2\Pi M s^2$ (*i.e.*, the theoretical value of SA) for both fcc-Py and a-Co films are plotted together. The numbers inside the plots represent the thickness in nm.

this, we understand that the magnetization in such films favours IP orientation more than OOP orientation. We name this 'mostly in-plane' (MIP) regime. The loop shape is more conventional "pot-belly" shaped in this regime, which is typical for hard magnets like Co. The transcritical shape is a salient feature for a-Co films with $d \geq 100$ nm in our experiment. This is indicated as ST regime here onwards for this material as well. [40][41]

Figure 4a-b show the normalized initial hysteresis curves for fcc-Py and a-Co films for both IP and OOP directions. The difference in the shape of the curves and their evolution is consistent with that of the entire hysteresis loops as described in Figure 3. Figure 4c-f contains the variation of $H_s$ and $K_{eff}$ with varying thicknesses of the films. The $K_{eff}$ is calculated from the difference between the area under the IP curve and the OOP curve for a sample [see Figures 7S and 8S]. For fcc-Py, $M_s$ = 816 emu/cc is obtained from the VSM measurements. For the RT regime of fcc-Py, $H_s$ (IP) ≈ $H_s$ (OOP). However, for the ST regime, $H_s$ (IP) < $H_s$ (OOP). Effective anisotropy shows a non-monotonic variation with increasing thickness. In this experiment, the major contributor of IP anisotropy is SA, which can be estimated from the relationship: $2\Pi M_s^2$.

It is reported that this anisotropy (SA) is independent of the film thickness and depends on the effective magnetization of the material. This arises due to the magnetic dipolar interactions. These interactions have a long-range nature and hence sense the boundaries of the specimen. It depends on the shape of the specimen. It forces magnetization to lie in-plane to minimize the demagnetizing field energy. In the absence of any other IP (and/or OOP) anisotropy, the $K_{eff}$ should be maximum when estimated from the area method (*i.e.*, area under OOP curve subtracted from area under IP curve). However, these films are deposited in rf sputtering with $\alpha \approx 45°$ (Figure 1a). When the vapour atoms arrive at the substrate surface, two processes might occur: either they can be reflected or be adsorbed by losing kinetic energy *via* lattice vibration. If adsorbed, the ad-atoms can bond with one another while forming a nucleus. Incoming ad-atoms may join the former one or may form another nucleus. They also may reflect to the opposite direction. However, this is less probable. Eventually, island-like features form following the Volmer-Weber growth model.[42][43] However, the increase in resistivity observed as thickness decreases cannot be attributed solely to the thickness of the film. While it is true that films deposited in off-normal condition have greater porosity than corresponding films grown at normal deposition, the increase in resistivity in thinner samples could be due to contribution from the defects. Comparing resistivity for normal and oblique angle (off-normal) deposition is beyond the scope of this work. The atoms approaching the substrate sit over the islands following the direction of the flux. This results in a shadowing effect, causing column-like structures to form. These canted column-like structures can have different shapes depending on specific deposition geometry and chamber conditions. Electrical resistivity is measured in the Van der Pauw geometry by using a custom-built four-probe setup. An increase in resistivity values for thinner films might indicate porosity introduced due to columnar growth [see supplementary information].[44][45] The magnetization will point towards the long axis of the column. This may give rise to a strong shape anisotropy, which we describe as growth-induced shape anisotropy (GI-AS). Typically, this supports OOP configuration for the magnetization. In addition to these two anisotropies, crystalline materials have MCA, which is the case for fcc-Py as suggested from the XRD spectra (see Figure 1b). The preferred orientation for the magnetization under this anisotropy will be <111>. [1] While discussing GI anisotropy and MCA, it is pertinent to mention here that MsA, which emerges from the same origin, could be present in such films. The films can be subjected to internal stress due to growth conditions and lattice mismatch between the film and the substrate. This stress can significantly modify the magnetization orientation for the magnetostrictive material. The magnetization direction will be determined by the combined effect of the sign of the stress ($\sigma$) and that of the magnetostriction constant ($\lambda$).[1] The nature of the GI-stress (planar tensile or planar compressive) is decided by a few major technical factors: deposition power, Ar pressure, deposition geometry, and heat treatment. On application of high deposition power, tensile stress may be developed inside the film. However, it can lead to compressive stress at lower power.[15] If the Ar pressure is maintained at a low value, then compressive stress arises in the films, which is opposite to the higher pressure case.[16] Formation of dense columnar structures will lead to compressive stress, and under-dense structures will lead to tensile stress, depending on the deposition geometry.[46] The detailed discussion about the effect of heat treatment can be found elsewhere. Physically, tension and compression in sputter-deposited films are governed by the energetics of the bombarding particles.[47] Porous- and sub-bulk deposition involving incoming flux with low energetic particles introduces a planar tensile stress. On the other hand, atomic pinning mediated by high-energy particles, hence the dense packing, leads to a planar compressive stress.[48] For crystalline materials, the straining of bonds between the substrate and thin layer leads to some residual stress inside the system. The lattice mismatch estimated from XRD data gives us an idea of having tensile stress for our samples. The magnetization prefers the direction of stress if $\lambda\sigma > 0$. The competition between all these individual anisotropies leads to an effective anisotropy that will ultimately decide the easy and hard directions for the magnetization in a film. This $K_{eff}$, due to the combined contribution of these anisotropies, can be written as:

$K_{eff} = K_{SA} + K_{GI-SA} + K_{MCA} + K_{MsA} + \ldots$

For our fcc-Py in the RT regime, $K_{eff}$ is low. We believe that SA points towards IP, GI-SA points OOP direction (along the long axis of the columns), and MCA points in <111> direction. The nature of $\lambda$ in the case of fcc-Py is identified from the composition variation with the thickness of the films [see supplementary information]. Depending on the chamber condition, it can be anticipated that a planar compressive stress is introduced, and this compressive stress, combined with the positive $\lambda$ of this set of Py films, forces magnetization to lie out of the sample plane. On the other hand, tensile stress produced due to the observed lattice mismatch in our fcc-Py films, with a positive value of $\lambda$, favours IP orientation. However, the strength of $K_{MCA}$ and $K_{MsA}$ will be insignificant at this thickness range due to obvious reasons. The competition between SA and GI-SA makes the magnetization point $\approx 35°$ from the IP direction in the demagnetized state. Thus, IP and OOP configurations in which hysteresis loops are obtained during the VSM measurement become the moderate axes towards which the magnetization can align itself with a similar preference. For both these configurations, the external magnetic field must provide almost an equal amount of energy to saturate the magnetization. Similarity in magnetization reversal nature and $H_s$ result in merging of the loops. Thus, the anisotropy suffers from underestimation when calculated from the area method, hence we have relatively low $K_{eff}$ for 5 nm $\leq d \leq$ 25 nm.

The conventional SA is independent of thickness, and it appears to be the major contributor in the ST regime for fcc-Py films. We anticipate that the contribution from GI-SA weakens because of the decreased shadowing effect due to substrate rotation.[49] We observe that the energy expense for the IP configuration is less than that of the OOP configuration. It implies dominance of IP anisotropy. However, persistence of weak GI-SA and emergence of OOP MCA help the films to accommodate flux closure domains and weak stripe domain-like features.[50] The easy axis might lie slightly towards the OOP direction, though we presume that the tilt will be less than that in the RT regime. It is typically observed that the stress-induced anisotropies become stronger with film thickness. For this set of films with negative $\lambda$, lattice mismatch-induced tensile stress results in OOP anisotropy. On the other hand, the compressive nature of growth-induced stress results in IP anisotropy. Combining them, the effect of stress is negligible for thicker Py films. However, the possibility of stress redistribution in this thickness regime cannot be ruled out. The anisotropy value in this regime corroborates some of the existing literature.

For a-Co films in their MIP regime, SA is the major contributor as usual. These films have possibly columnar growth and hence GI-SA, like the fcc-Py films. However, there are no other contributors in the OOP direction that overpower the SA. Thus, the magnetization lies close to the plane of the films between 5 nm $\leq t \leq$ 90 nm. Moving forward to the ST regime, the OOP stress-induced anisotropy is anticipated to grow with film thickness. Beyond a critical thickness ($t \geq$ 100 nm), initiation of stripe domain is evident from the so called "transcritical" shape of the hysteresis loops and from the reduction in $K_{eff}$ value. These kind of loops arise near spin-reorientation transition (SRT) wherein, shape anisotropy competes with OOP component of magnetic anisotropy.[51] The reduction in $K_{eff}$ can be attributed to the competing anisotropies.

In essence, the fcc-Py and a-Co films suffer from the effect of mixed anisotropies. As a result, none of the films exhibit conventional magnetization reversal with low coercivity, high retentivity, and a rectangular-shaped hysteresis loop in the IP configuration. Instead of that, the fcc-Py films in the experimental thickness range ($d$ = 5-125 nm) are categorized in two ways: in RT regime, magnetization is believed to have more OOP inclination than that of the ST regime in a demagnetized state. On the contrary, thinner a-Co films seem to prefer IP orientation more than the fcc-Py films. Thus, initiation of the patterned domain structures occurs at a higher thickness range in a-Co than fcc-Py. The preferred state of magnetization and possible anisotropy directions in different regimes are pictorially represented in Figure 5. This is a humble attempt to create a visual impression that provides better correlation between our experimental results and their explanations.

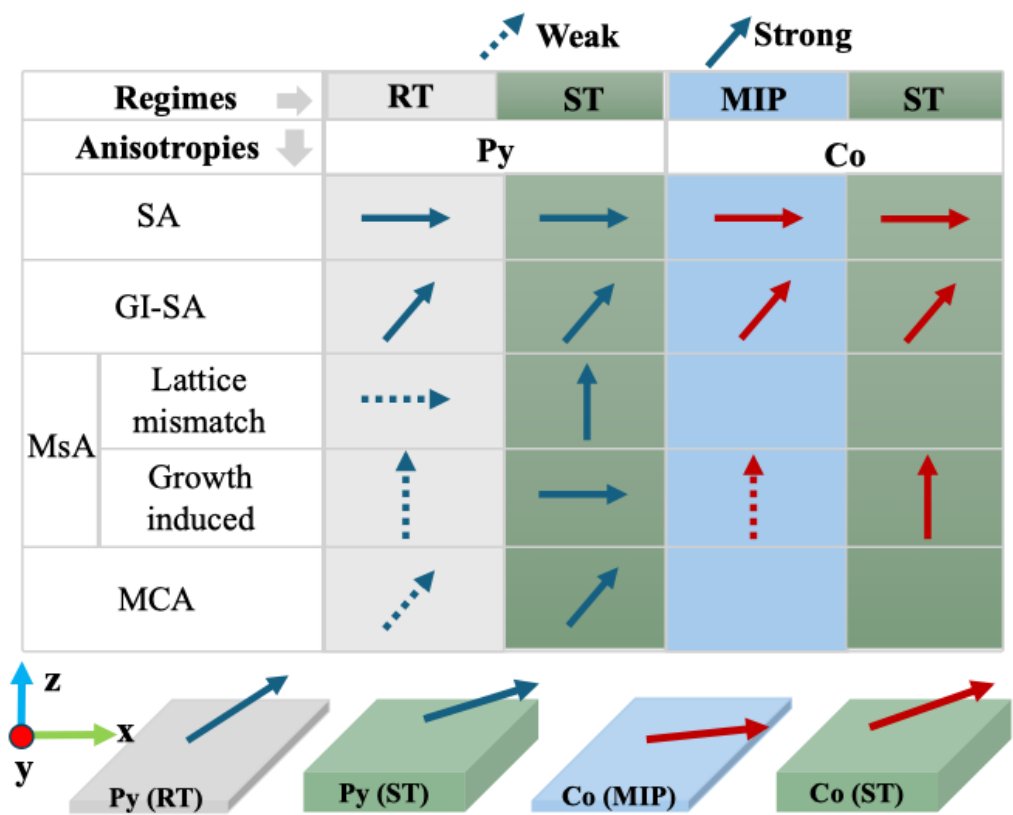

FIGURE 5: Schematic diagram of individual anisotropy contributions and their cumulative effect on magnetization orientation of thin films in the demagnetized state for both fcc-Py and a-Co films. The colour, length, and orientations of the arrows do not represent quantitative changes. Here, SA, GI-SA, MsA and MCA refer to the shape anisotropy, growth induced- shape anisotropy, magnetostrictive anisotropy and magnetocrystalline anisotropy respectively.

## 4 Conclusions

In summary, we investigate the origin of mixed anisotropies within a broad thickness range for sputter-deposited fcc-Py and a-Co thin films. We correlate those findings with structural, electrical, and growth-induced properties, such as crystallinity, roughness, resistivity, possible formation of columnar structure, magnetostrictions, stress, etc. We have observed that modification of magnetic hysteresis behaviour arises due to the competition between several anisotropy components. The presence of shape anisotropy, growth-induced shape anisotropy, magnetostrictive anisotropy (due to lattice mismatch and growth-induced stress), and magnetocrystalline anisotropy is predicted from the analysis of the hysteresis curves. It is observed that the shape anisotropy, being the major contributor in fixing the magnetization IP, must face a strong competition from growth-induced shape anisotropy. The role of magnetostrictive anisotropy and magnetocrystalline anisotropy in modifying the magnetization states becomes substantial for the relatively thicker samples. The fcc-Py films behave distinctly in the robustly tilted (RT) and subtly tilted (ST) regimes. However, for a-Co films, this anomalous RT regime is not present. Instead of that, more conventional mostly in-plane (MIP) and ST regimes are clearly identifiable due to the absence of two anisotropic competitors, lattice mismatch-induced magnetostrictive and magnetocrystalline, unlike fcc-Py. Later, by analysing the shape of the hysteresis curves, we presume that the interplay of these mixed anisotropies has led not only to the change of the overall magnetization tilt in the demagnetized state but also to the formation of complex domain structures during the reversal. It is pertinent to mention that the efficiency of any magnetic device is governed by the structural compatibility and magnetic tunability of the material embedded into it. Introduction of magnetization tilt has been proven to be an extremely innovative way to enhance performance quality for such devices. Our investigation of anisotropic behaviour in sputter-deposited crystalline and amorphous samples might help in building a concrete understanding while designing device prototypes in the future.

## 5 Experimental Techniques

(a) Sample deposition by using rf-magnetron sputtering: Thin films of Py and Co with thicknesses $d$ = 5-125 nm and $t$ = 5-150 nm are deposited at room temperature on p-type Si (100) substrate by using a high-vacuum rf magnetron sputtering (designed by Excel instruments). The base pressure of the chamber was maintained below $7 \times 10^{-7}$ Torr. Both Py and Co were deposited at an Ar working pressure of 7 mTorr using rf source power of 80 W at 13.56MHz, and Ar flow rate of 40 SCCM. The substrate to gun distance is about 7 cm. To ensure uniformity, all the deposition conditions are carefully optimized and kept identical across all the samples of interest. The deposition flux approached the substrate at an angle of 45° relative to the substrate normal, *i.e.*, referred to as the incidence angle, as shown in Figure 1a, while the substrate is rotated at 2 rpm. Each sample has lateral dimensions of $1 \times 1$ cm$^2$.

(b) Thickness measurement by using profilometry: The Bruker Dektak profilometer system was used to estimate the thickness of all samples exceeding 50 nm, allowing for the optimization of their thickness, which indicates a deposition rate of approximately 0.8 Å/sec. [see supplementary information, Figure 1S].

(c) Investigating crystalline properties using XRD technique: The crystallinity of Py and Co films is investigated from GI-XRD measurements by using Cu K$_α$ (0.154187 nm) radiation on a Rigaku Smartlab SE diffractometer operated at 3kW. All measurements were performed at room temperature within the range $2\theta = 40°$ to $55°$, except for the measurement of the Si substrate in bulk mode at a scan rate of 10°/min with 0.02° resolution. The background-subtracted XRD spectra for the Si substrate, Py, and Co samples are presented in Figure 1b-c. Peak intensities with respect to the baseline are retained as obtained during the XRD measurement in these plots.

(d) Verifying the surface quality with AFM technique: To probe the surface topography, AFM measurements were conducted in intermittent contact (tapping) mode using Oxford Instruments MFP-3D ORIGIN Asylum Research by taking a scan over a $5~\mu m \times 5~\mu m$ area for all the samples except one Co film ($t$ = 5 nm, ~ $4~\mu m \times 4~\mu m$), and analysed using Gwyddion software.

(e) Investigating hysteresis behaviour using VSM: The static magnetic properties are studied by using VSM attached to a Quantum Design Dynacool PPMS instrument. The external magnetic field is swept between 2T to -2T with 20 Oe field resolution near the reversal. The measurement was performed at room temperature for all the samples. The hysteresis loops are obtained in both parallel and perpendicular directions with the magnetic field.

**Acknowledgements**

We acknowledge the financial support from Shiv Nadar Institution of Eminence (Delhi NCR) and INSPIRE fellowship (Department of Science and Technology, Govt. of India, grant number: G-CONF000220). K.K., B.G., and A.M. acknowledge Shiv Nadar Institution of Eminence (Delhi NCR) for research fellowships. R. K. acknowledges the INSPIRE scheme (Department of Science and Technology, Govt. of India, grant number: G-CONF000220) for the stipend during his project assistantship period. We would also like to acknowledge the financial support from the Department of Science and Technology, India, under the FIST project (grant number: SR/FST/PS-I/2017/6(C)) for the installation and maintenance of a custom-built Hall measurement setup. We are grateful to our scientific officers, Dr. Rakesh Kumar and Dr. Khushboo Agarwal from the Department of Physics (Shiv Nadar Institution of Eminence, Delhi NCR), for their guidance in handling PPMS, sputtering technique, and resistivity measurements. We are glad to acknowledge technical support from Mr. Ravinder Singh and Dr. Aniruddha Das from the Department of Chemistry (Shiv Nadar Institution of Eminence, Delhi NCR) for AFM and XRD (Rigaku X-ray diffractometer 3kW system, model Smartlab SE funded by Shiv Nadar Foundation) measurements, respectively. We also acknowledge Dr. Raju Vemoori from the Department of Mechanical Engineering (Shiv Nadar Institution of Eminence, Delhi NCR) for helping us obtain EDX data. We are thankful to Miss Sukanaya Handique and Harsh Munout for their help during their OUR project.

**Author Contribution**

Kirti Kirti (Conceptualization: Equal; Data curation: Lead; Formal analysis: Lead; Investigation: Equal; Methodology: Lead; Software: Lead; Validation: Equal; Visualization: Equal; Writing – original draft: Equal; Writing – review and editing: Equal), Baisali Ghadai (Conceptualization: Supporting; Data curation: Equal; Formal analysis: Equal; Methodology: Equal: Visualization: Equal; Writing – original draft: Supporting; Writing – review and editing: Supporting), Abinash Mishra (Data curation: Supporting; Methodology: Supporting; Writing – original draft: Supporting; Writing – review and editing: Supporting), Rahulkrishnan R (Data curation: Supporting; Formal analysis: Supporting), Sucheta Mondal (Conceptualization: Lead; Funding acquisition: Lead; Project administration: Lead; Resources: Lead; Software: Lead; Supervision: Lead; Validation: Lead; Visualization: Equal; Writing – original draft: Equal; Writing – review and editing: Equal)

**Conflict of interest**

The authors have declared no conflict of interest.

**Keywords**

Ferromagnet, Thin film, Magnetic hysteresis, Mixed anisotropy, Vibrating sample magnetometry

**References**

1. Cullity, B.D., and Graham, C.D. (2011) *Introduction to magnetic materials*, John Wiley & Sons.

2. Jensen, P.J., and Bennemann, K.H. (2006) Magnetic structure of films: Dependence on anisotropy and atomic morphology. *Surf. Sci. Rep.*, **61** (3), 129–199.

3. Delprat, S., Galbiati, M., Tatay, S., Quinard, B., Barraud, C., Petroff, F., Seneor, P., and Mattana, R. (2018) Molecular spintronics: The role of spin-dependent hybridization. *J. Phys. D. Appl. Phys.*, **51** (47).

4. A. Fert, and Peter M Levy (1980) Role of Anisotropic Exchange Interactions in Determining the Propertiesof Spin-Glasses. *Phys. Rev. Lett.*, **44** (23), 1538–1541.

5. Parkin, S.S.P., More, N., and Roche, K.P. (1990) Oscillations in exchange coupling and magnetoresistance in metallic superlattice structures: Co/Ru, Co/Cr, and Fe/Cr. *Phys. Rev. Lett.*, **64** (19), 2304–2307.

6. Meiklejohn, W.H., and Bean, C.P. (1956) New magnetic anisotropy [1]. *Phys. Rev.*, **102** (5), 1413–1414.

7. Williams, C.M., and Schindler, A.I. (1966) Composition dependence of irradiation-induced uniaxial anisotropy energy of Permalloy films. *J. Appl. Phys.*, **37** (3), 1468–1469.

8. Gubbiotti, G., Tacchi, S., Madami, M., Carlotti, G., Adeyeye, A.O., and Kostylev, M. (2010) Brillouin light scattering studies of planar metallic magnonic crystals. *J. Phys. D. Appl. Phys.*, **43** (26).

9. Rottmayer, R.E., Batra, S., Buechel, D., Challener, W.A., Hohlfeld, J., Kubota, Y., Li, L., Lu, B., Mihalcea, C., Mountfield, K., Pelhos, K., Peng, C., Rausch, T., Seigler, M.A., Weller, D., and Yang, X. (2006) Heat-Assisted Magnetic Recording. **42** (10), 2417–2421.

10. Fassbender, J., Von Borany, J., Mücklich, A., Potzger, K., Möller, W., McCord, J., Schultz, L., and Mattheis, R. (2006) Structural and magnetic modifications of Cr-implanted Permalloy. *Phys. Rev. B - Condens. Matter Mater. Phys.*, **73** (18), 1–8.

11. Journal of the Physical Society of Japan 1955 Chikazumi S.pdf.

12. Takahashi, M. (1962) Induced magnetic anisotropy of evaporated films formed in a magnetic field. *J. Appl. Phys.*, **33** (3), 1101–1106.

13. Craik, D.J., and Wood, M.J. (1970) Magnetization changes induced by stress in a constant applied field. *J. Phys. D. Appl. Phys.*, **3** (7), 1009–1016.

14. Weisheit, M., Fähler, S., Marty, A., Souche, Y., Poinsignon, C., and Givord, D. (2007) Electric field-induced modification of magnetism in thin-film ferromagnets. *Science (80-. ).*, **315** (5810), 349–351.

15. Cheng, S.F., Lubitz, P., Zheng, Y., and Edelstein, A.S. (2004) Effects of spacer layer on growth, stress and magnetic properties of sputtered permalloy film. *J. Magn. Magn. Mater.*, **282** (1–3), 109–114.

16. Kurlyandskaya, G. V (2010) Near the " Transcritical " State. **46** (2), 333–336.

17. Kim, J.S., Koo, Y.M., Lee, B.J., and Lee, S.R. (2006) The origin of (001) texture evolution in FePt thin films on amorphous substrates. *J. Appl. Phys.*, **99** (5).

18. M T Johnson, P J H Bloemen, F J A den Broeder, J.J.V. (1996) Magnetic anisotropy in metallic multilayers. *Rep. Prog. Phys.*, **64** (111), 297–381.

19. Kim, J. Von, Demand, M., Hehn, M., Ounadjela, K., and Stamps, R.L. (2000) Roughness-induced instability in stripe domain patterns. *Phys. Rev. B - Condens. Matter Mater. Phys.*, **62** (10), 6467–6474.

20. Beach, G.S.D., Knutson, C., Nistor, C., Tsoi, M., and Erskine, J.L. (2006) Nonlinear domain-wall velocity enhancement by spin-polarized electric current. *Phys. Rev. Lett.*, **97** (5), 8–11.

21. Roy, P.E. (2013) In-plane anisotropy control of the magnetic vortex gyrotropic mode. *Appl. Phys. Lett.*, **102** (16).

22. Kent, N., Reynolds, N., Raftrey, D., Campbell, I.T.G., Virasawmy, S., Dhuey, S., Chopdekar, R. V., Hierro-Rodriguez, A., Sorrentino, A., Pereiro, E., Ferrer, S., Hellman, F., Sutcliffe, P., and Fischer, P. (2021) Creation and observation of Hopfions in magnetic multilayer systems. *Nat. Commun.*, **12** (1), 1–7.

23. Mühlbauer, S. (2011) Skyrmion lattice in a chiral magnet (Science (2009) (915)).

*Science (80-. ).*, **333** (6048), 1381.

24. Streubel, R., Han, L., Im, M.Y., Kronast, F., Rößler, U.K., Radu, F., Abrudan, R., Lin, G., Schmidt, O.G., Fischer, P., and Makarov, D. (2015) Manipulating topological states by imprinting non-collinear spin textures. *Sci. Rep.*, **5**, 1–7.

25. Demokritov, S.O., Hillebrands, B., and Slavin, A.N. (2001) Brillouin light scattering studies of confined spin waves: Linear and nonlinear confinement. *Phys. Rep.*, **348** (6), 441–489.

26. Kittel, C. (1948) On the theory of ferromagnetic resonance absorption. *Phys. Rev.*, **73** (2), 155–161.

27. Chen, L., Mankovsky, S., Wimmer, S., Schoen, M.A.W., Körner, H.S., Kronseder, M., Schuh, D., Bougeard, D., Ebert, H., Weiss, D., and Back, C.H. (2018) Emergence of anisotropic Gilbert damping in ultrathin Fe layers on GaAs(001). *Nat. Phys.*, **14** (5), 490–494.

28. Mizukami, S., Wu, F., Sakuma, A., Walowski, J., Watanabe, D., Kubota, T., Zhang, X., Naganuma, H., Oogane, M., Ando, Y., and Miyazaki, T. (2011) Long-lived ultrafast spin precession in manganese alloys films with a large perpendicular magnetic anisotropy. *Phys. Rev. Lett.*, **106** (11), 1–4.

29. Hämäläinen, S.J., Madami, M., Qin, H., Gubbiotti, G., and van Dijken, S. (2018) Control of spin-wave transmission by a programmable domain wall. *Nat. Commun.*, **9** (1), 1–8.

30. Xiao, H.Z., Ye, F., Zhou, D.Y., and Bai, F.M. (2014) Structural transformation relationship for hafnia ferroelectric phase. *Adv. Mater. Res.*, **873** (96), 865–870.

31. Liu, L., Pai, C.F., Li, Y., Tseng, H.W., Ralph, D.C., and Buhrman, R.A. (2012) Spin-torque switching with the giant spin hall effect of tantalum. *Science (80-. ).*, **336** (6081), 555–558.

32. You, L., Lee, O.J., Bhowmik, D., Labanowski, D., Hong, J., Bokor, J., and Salahuddin, S. (2015) Switching of perpendicularly polarized nanomagnets with spin orbit torque without an external magnetic field by engineering a tilted anisotropy. *Proc. Natl. Acad. Sci. U. S. A.*, **112** (33), 10310–10315.

33. Stanciu, C.D., Hansteen, F., Kimel, A. V., Kirilyuk, A., Tsukamoto, A., Itoh, A., and

Rasing, T. (2007) All-optical magnetic recording with circularly polarized light. *Phys. Rev. Lett.*, **99** (4), 1–4.

34. Unikandanunni, V., Medapalli, R., Fullerton, E.E., Carva, K., Oppeneer, P.M., and Bonetti, S. (2021) Anisotropic ultrafast spin dynamics in epitaxial cobalt. *Appl. Phys. Lett.*, **118** (23).

35. Sahoo, A., Mahanta, S.P., and Bedanta, S. (2025) Significant influence of low SOC materials on magnetization dynamics and spin-orbital to charge conversion. *npj Spintron.*, **3** (1), 1–9.

36. Guittoum, A., Bourzami, A., Layadi, A., and Schmerber, G. (2012) Structural, electrical and magnetic properties of evaporated permalloy thin films: Effect of substrate and thickness. *EPJ Appl. Phys.*, **58** (2), 1–6.

37. Roth, T.A. (1975) The Surface and Grain Boundary Energies of Iron, Cobalt and Nickel. *Mater. Sci. Eng.*, **18**.

38. Mondal, S., and Barman, A. (2018) Laser Controlled Spin Dynamics of Ferromagnetic Thin Film from Femtosecond to Nanosecond Timescale. *Phys. Rev. Appl.*, **10** (5), 1.

39. Kharmouche, A. (2011) Thickness dependent magnetic and structural properties of Co xCr 1-x thin films evaporated on Si(100) and glass substrates. *J. Nanosci. Nanotechnol.*, **11** (6), 4757–4764.

40. Hehn, M., Padovani, S., Ounadjela, K., and Bucher, J. (1996) Nanoscale magnetic domain structures in epitaxial cobalt films. *Phys. Rev. B - Condens. Matter Mater. Phys.*, **54** (5), 3428–3433.

41. Lisfi, A., and Lodder, J.C. (2001) Magnetic domains in Co thin films obliquely sputtered on a polymer substrate. *Phys. Rev. B - Condens. Matter Mater. Phys.*, **63** (17), 1–5.

42. Ali, Z., Basaula, D., Eid, K.F., and Khan, M. (2021) Anisotropic properties of oblique angle deposited permalloy thin films. *Thin Solid Films*, **735** (August), 138899.

43. Barranco, A., Borras, A., Gonzalez-Elipe, A.R., and Palmero, A. (2016) Perspectives on oblique angle deposition of thin films: From fundamentals to devices. *Prog. Mater. Sci.*, **76**, 59–153.

44. Lapteva, N.N., and Kulikova, L.I. (1975) Nekotorye metodologicheskie aspekty kolichestvenno-kachestvennykh narusheniĭ obmena veshchestv. *Vestn. Akad. Med. Nauk SSSR*, (5), 76–78.

45. Kharmouche, A., and Cherrad, O. (2024) Electrical properties of permalloy/Si (100) thin films. *J. Mater. Sci. Mater. Electron.*, **35** (11), 1–12.

46. Abadias, G., Chason, E., Keckes, J., Sebastiani, M., Thompson, G.B., Barthel, E., Doll, G.L., Murray, C.E., Stoessel, C.H., and Martinu, L. (2018) Review Article: Stress in thin films and coatings: Current status, challenges, and prospects. *J. Vac. Sci. Technol. A Vacuum, Surfaces, Film.*, **36** (2).

47. Sriram, K., Mondal, R., Pradhan, J., Haldar, A., Murapaka, C. structural-phase-engineering-of-(α-β)-w-for-a-large-spin-hall-angle-and-spin-diffusion-length.pdf.

48. Thornton, J.A., Tabock, J., and Hoffman, D.W. (1979) Internal stresses in metallic films deposited by cylindrical magnetron sputtering. *Thin Solid Films*, **64** (1), 111–119.

49. Shim, Y., Mills, M.E., Borovikov, V., and Amar, J.G. (2009) Effects of substrate rotation in oblique-incidence metal(100) epitaxial growth. *Phys. Rev. E - Stat. Nonlinear, Soft Matter Phys.*, **79** (5), 1–6.

50. Singh, S., Gao, H., and Hartmann, U. (2018) Nucleation of stripe domains in thin ferromagnetic films. *Phys. Rev. B*, **98** (6), 1–6.

51. Xi, L., Lu, J.M., Zhou, J.J., Sun, Q.J., Xue, D.S., and Li, F.S. (2010) Thickness dependence of magnetic anisotropic properties of FeCoNd films. *J. Magn. Magn. Mater.*, **322** (15), 2272–2275.

# Supplementary Information

# Origin of mixed anisotropy in crystalline Permalloy and amorphous Cobalt thin films individually deposited on Si substrate

Kirti Kirti, Baisali Ghadai, Abinash Mishra, Rahulkrishnan R, Sucheta Mondal*

*Shiv Nadar Institution of Eminence, Greater Noida, Delhi NCR, UP-201314*

*sucheta.mondal@snu.edu.in

**Determination of film thickness and deposition rate**

Figure 1S represents the height profile obtained from a stylus-based profilometer [Bruker Dektak profilometer]. The measured thicknesses for all the Permalloy ($Ni_{80}Fe_{20}$, Py hereafter) and Cobalt (Co) films differ by ∼5% of the nominal thickness. The deposition rate (measured thickness/ measured time of deposition) is about 0.8 Å/sec. The thickness estimation for thinner samples (< 50 nm) is not accurately obtained from this technique because of the instrumental limitations. We have extrapolated those values for the low thickness range accordingly.

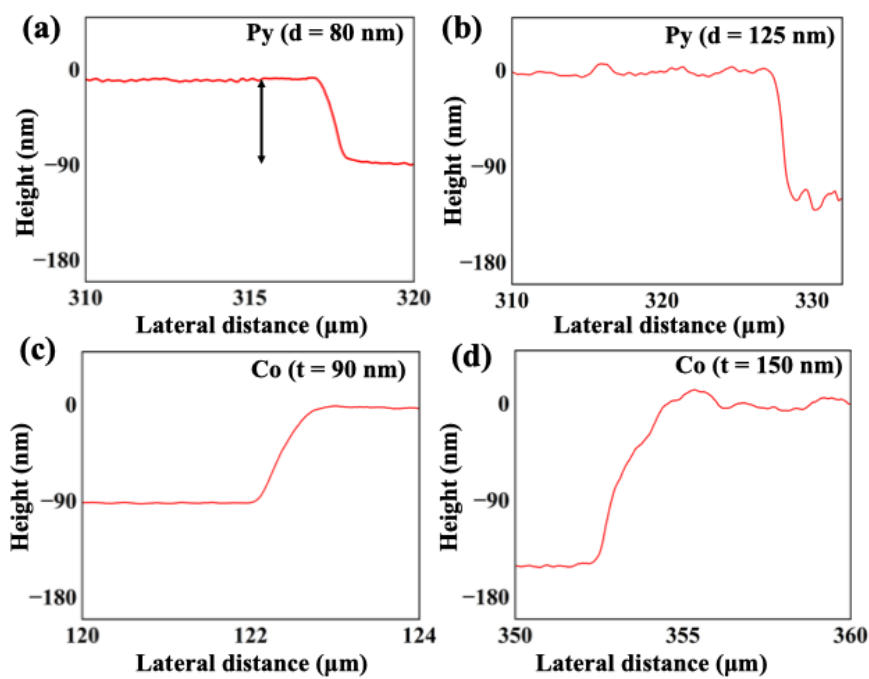

FIGURE 1S: Line profile of the (a-b) fcc-Py ($d$ = 80 and 125 nm), (c-d) a-Co ($t$ = 90 and 150 nm) films. The double-sided arrow represents the depth at which height is measured.

**Analysis of XRD spectra for substrate and samples**

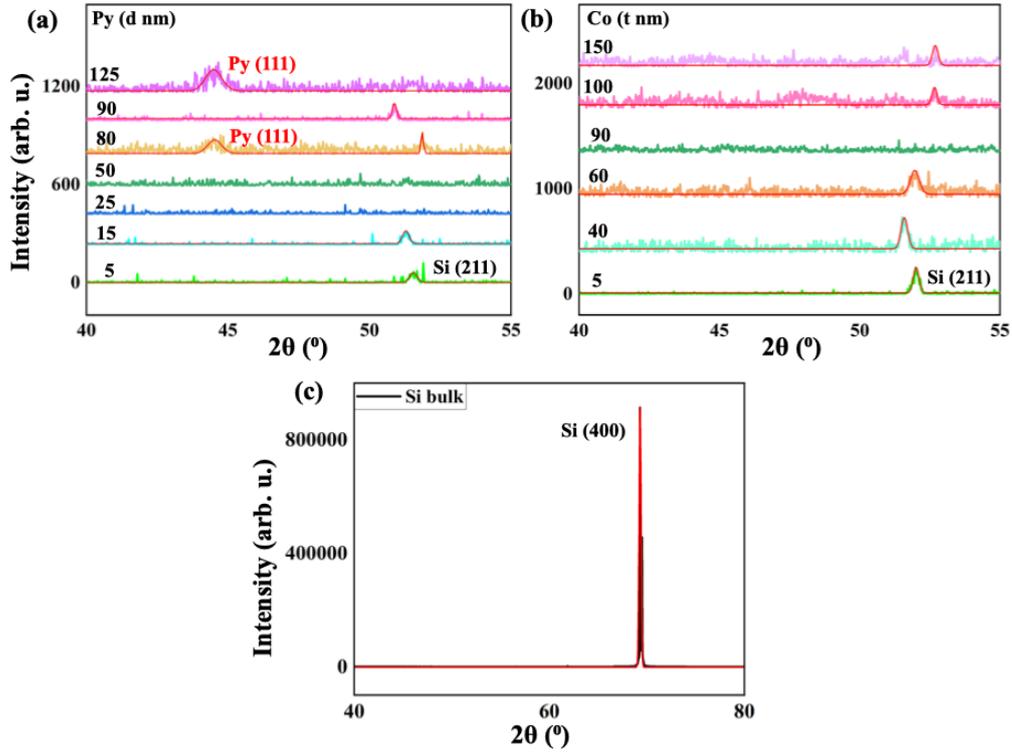

FIGURE 2S: XRD spectra for (a) fcc-Py and (b) a-Co films, respectively obtained in GI mode showing substrate peak at around 52°, (c) XRD spectra for the Si substrate obtained in bulk mode. The numbers represent the thickness in nm. The (hkl) values are mentioned for the individual peaks

Figure 2S represents the XRD spectra for all the Py and Co samples after fitting the desired peaks with a Gaussian function and extracting FWHM from the fitting we have applied, Debye-Scherrer formula as follows:

$$D = \frac{k\lambda}{\beta \cos\theta}$$

Where, $D$ is crystallite size, $k$ is shape factor (0.9), $\lambda$ is the x-ray wavelength, $\beta$ is the line broadening at half the maximum intensity (FWHM) and $\theta$ is the Bragg's angle.

The crystallite size is between 12 nm and 15nm for fcc-Py. Using Bragg's law

$$2d \sin\theta = n\lambda$$

where, $d$ is the inter-planar spacing, $\theta$ is the Bragg's angle, $n$ is the order diffraction, and $\lambda$ is the wavelength of the X-ray. The lattice constants for the fcc-Py and Si are found to be 0.35 nm and 0.54 nm, respectively. Using the following formula $\eta = \frac{(a_f - a_s)}{a_f}$ here, $\eta$ is lattice mismatch.[18] Here, $a_f$ and $a_s$ are the lattice constants of the material and the substrate, respectively. We estimate the lattice mismatch to be -54%.

**AFM image of the substrate**

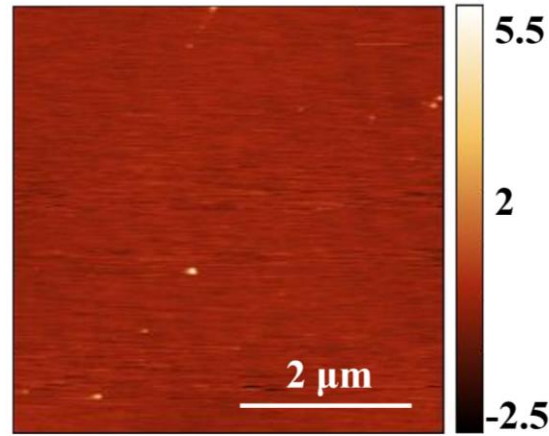

FIGURE 3S: AFM image of the Si substrate processed using Gwyddion software. The scan area presented in the figures is 5 $\mu m \times 5\ \mu m$ . Length scale is provided for reference. The colour bar indicates the height profile in nm.

**Oblique angle deposition and effect of substrate rotation**

The oblique-angle deposition can lead to column-like structures in thin films.[41] The canting angle ($\beta$) of these structures depends strongly on the angle of deposition ($\alpha$): $tan(\alpha) = 2tan(\beta)$.[42] Since for our deposition geometry, $\alpha \approx 45^o$, we calculate $\beta$ for our films to be $\approx 22^o$ with film-normal [see Figure 4S]. This columnar structure, leading to anisotropy in the long axis of the columns, is termed growth-induced shape anisotropy (GI-SA).

However, in our case, the substrate holder is slowly rotated, which causes a reduction in the shadowing effect and affects this columnar growth. Thus, a decrease in the anisotropy strength might occur at higher thicknesses.

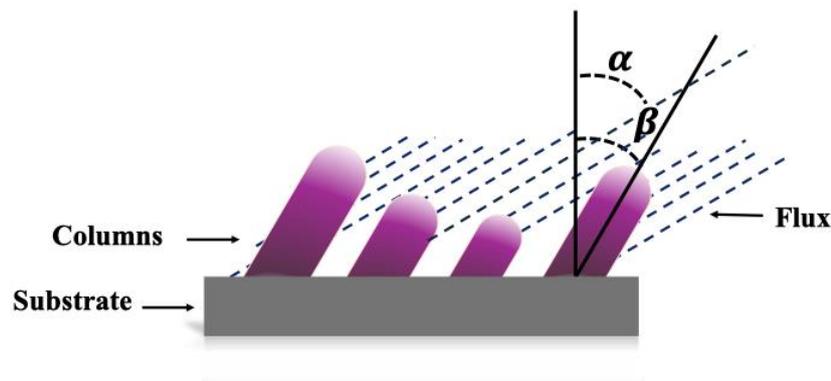

FIGURE 4S: Artist's impression of the possible columnar growth created during oblique-angle deposition. α and β, are the angles of incident flux and the orientation of the columns with respect to the substrate normal, respectively.

**Quantitative analysis of hysteresis for magnetic samples**

Figure 5S shows normalized in-plane (IP) and out-of-plane (OOP) hysteresis loops for the magnetic field ranging between 2 T and -2 T. The IP hysteresis loops exhibit delayed saturation beyond 95% remanence. We suspect the presence of magnetic inhomogeneities and structural defects within our films, which might create local pinning centres for magnetization to stay OOP. We believe a fraction of such stubborn domains need more than 8000 Oe field to orient them in the plane of the film, which contributes to 2-5% of the loss in retentivity on a case-by-case basis. For fcc-Py ($d$ = 125 nm) film, low signal-to-noise ratio of the loop indicates the presence of possible oxide formation the surface, which might have masked the transcritical loop shape in comparison to the other films of ST (subtly tilted) regime.

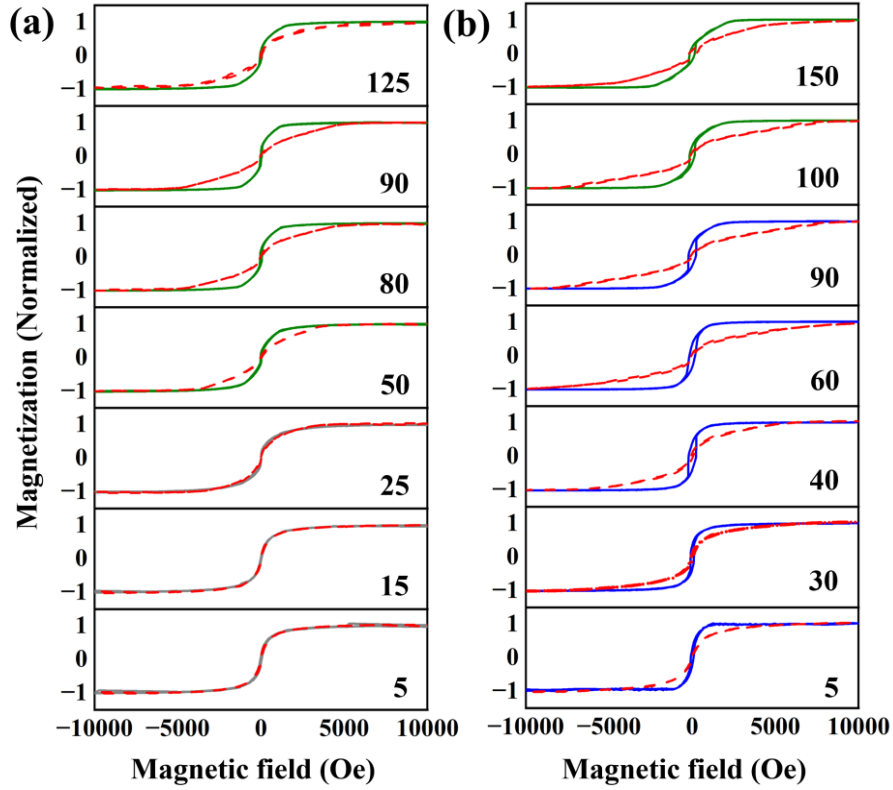

FIGURE 5S: Normalized IP (solid line) and OOP (dashed line) hysteresis loops for the (a) fcc-Py and (b) a-Co films, respectively, in the field range of ± 1T. Different regimes are presented in different colours: for (a) RT (5 nm ≤ $d$ ≤ 25 nm) in grey and ST (50 nm ≤ $d$ ≤ 125 nm) in olive; for (b) MIP (5 nm ≤ $t$ ≤ 90 $nm$) in blue and ST (100 nm ≤ $t$ ≤ 150 $nm$) in olive. The numbers inside the plots represent the thickness in nm.

Figure 6S shows the magnetic parameters obtained from hysteresis loops for both fcc-Py and a-Co films. We observe from the $H_c$ plot for fcc-Py, the value increases as we move towards the transcritical shape regime. A similar observation can be made from the $H_c$ plots for a-Co samples.[1S] The same trend is also followed by $H_s$ for both samples. However, the overlap of $H_s$ (IP) and $H_s$ (OOP) values can be seen for fcc-Py samples in the RT (robustly tilted) regime, which suggests a similar preference of magnetization in both configurations. The absence of a rectangular-shaped loop suggests that these directions are rather moderate axes. In the ST, regime, the gap between the $H_s$ (IP) and $H_s$ (OOP) is increasing, which suggests the dominance of one type of configuration over the other. In case of a-Co, no such overlap can be seen, indicating the absence of RT regime. The $H_s$ values are increasing with film thickness, but $H_s$ (IP) is always smaller than $H_s$ (OOP), suggesting IP configuration is preferred over the OOP in most of the films. This regime is named as MIP (mostly in-plane) regime in the main manuscript. As we move towards the ST regime, there is a decrease in and $H_s$ (OOP), indicating the emergence of the OOP component. The remanence ratio, $M_r/M_s$ (%) values, also have similar overlap in the RT regime for the fcc-Py films. This supports the claim that the magnetization prefers to orient itself along both directions with similar importance. Whereas high IP $M_r/M_s$ (%) can be seen in the ST regime, suggesting the dominance of IP preference, but a slight increase in OOP $M_r/M_s$ (%) can be linked to the emergence of the OOP component. In the case of a-Co IP, $M_r/M_s$ (%) is always higher than that of OOP, indicating a more coherent rotation of the magnetization domains during reversal. However, a shift in the values, an increase of OOP, $M_r/M_s$ (%) and a decrease of IP $M_r/M_s$ (%) indicate the introduction OOP component.

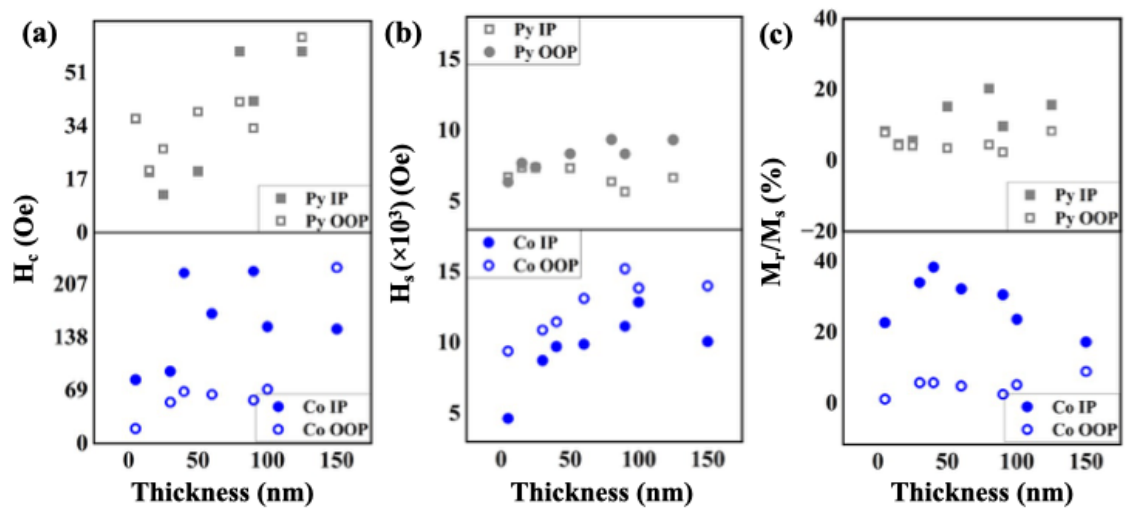

FIGURE 6S: (a) variation of coercivity ($H_c$), (b) saturation field ($H_s$), and remanence ratio: $M_r/M_s$ (%) for both fcc-Py and a-Co films obtained from the hysteresis loops.

Figure 7S shows that the variation of saturation magnetization with sample thickness is almost independent of the film thickness for both fcc-Py and a-Co. Due to the technical limitation of the instrument, the total magnetic moment for the thinner films could not be obtained with reliability. $M_s$ = 816 and 1236 emu/cc are considered for fcc-Py and a-Co, respectively, throughout the manuscript.

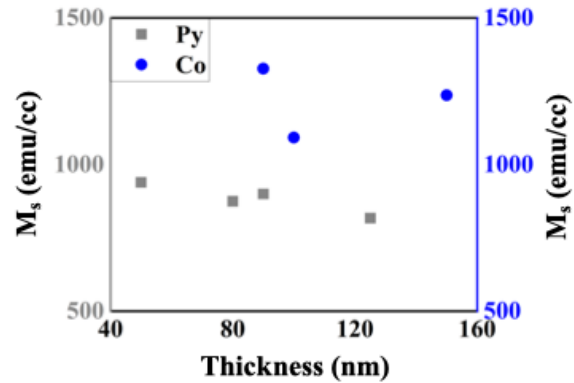

FIGURE 7S: $M_s$ vs thickness plot for fcc-Py and a-Co films.

Figure 8S explains the process of estimating effective anisotropy ($K_{eff}$) by subtracting the area under the normalized initial curve measured in the IP and OOP configurations. The resulting value is multiplied by a factor $M_s$ to get anisotropy in cgs units.

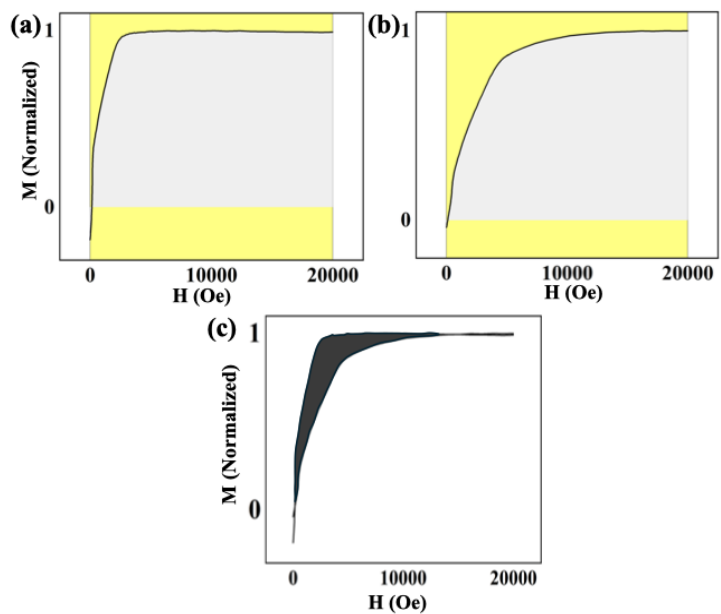

FIGURE 8S: Illustration of the area method. (a-b) The area under the curve for IP and OOP initial curves, respectively, (c) area (IP) – area (OOP) denoted with the shaded region. The data for a-Co ($t$ = 50 nm) sample is used for this demonstration.

**Electrical characterization and estimation of resistivity**

Figure 9S contains the information about the variation of resistivity with thickness for fcc-Py and a-Co samples, measured in the Van der Pauw geometry by using a four-probe setup. The values confirm the metallic nature of the films. An increase in resistivity of thinner samples might passively indicate the porous nature of the layer, which may originate from chamber conditions. The resistivity values are not out of range, which suggests that films are mostly metallic and continuous.

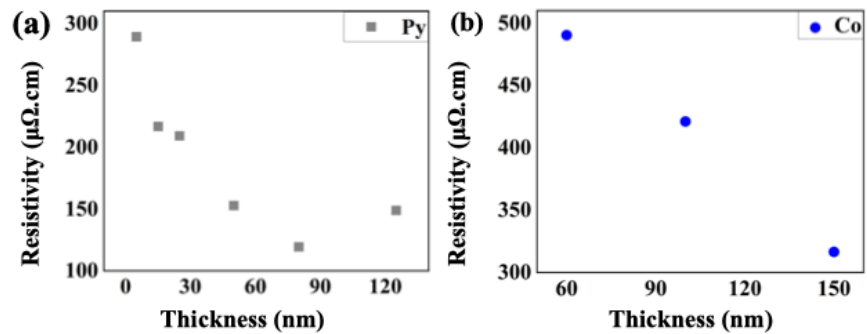

FIGURE 9S: The variation of resistivity with thickness for (a) fcc-Py and (b) a-Co films.

**Elemental analysis by using energy-dispersive X-ray analysis and correlation with magnetostriction values**

Figure 10S shows the composition variation for fcc-Py and a-Co with respect to thickness. In the case of Py, it can be seen that the Fe % has changed with thickness. We believe that the chamber conditions have led to this change in composition. As it has been studied previously, a change in deposition parameters can lead to a change in Fe content in Py films.[16] We believe that during deposition, some change in chamber conditions has resulted in this. The reason behind this has not been figured out.

The change in Fe and Ni composition can modify the magnetostriction of Py.[2S] The variation of the sign of the magnetostriction constant ($\lambda$) with respect to the change in Ni composition is presented in the following table.

TABLE 1S: represent the magnetostriction constant sign with the change in Ni %.

| Ni weightage | <80 | ≈80 | >80 |
|---|---|---|---|
| Sign of $\lambda$ | Positive | Zero | Negative |

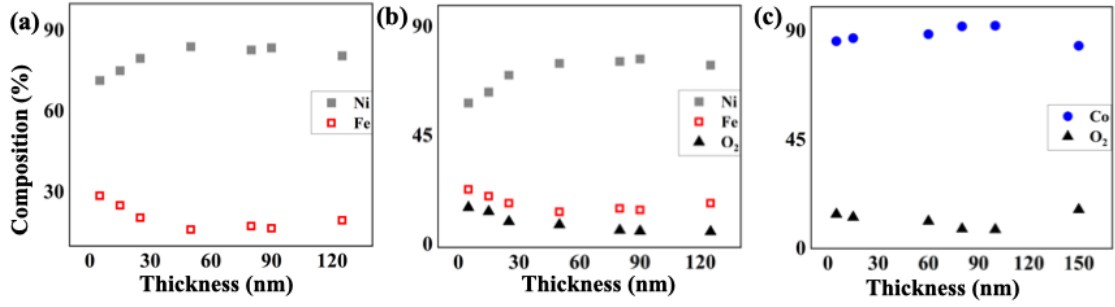

FIGURE 10S: (a) The change in elemental composition of Ni and Fe in fcc-Py films, (b-c) the $O_2$ content in the film for fcc-Py and a-Co, respectively.

Figure 11S represents how the nature of stress ($\sigma$) and $\lambda$ affects the effective magnetoelastic anisotropy (magnetostrictive anisotropy, MsA) due to stress arising from lattice mismatch and growth conditions. As established from the XRD data that we have a lattice mismatch of -54%. This lattice mismatch induces a tensile strain as $\varepsilon_f \propto -\eta$. Here, $\varepsilon_f$ is the strain induced in the film due to the lattice mismatch, $\eta$ between the film and the substrate. The stress-induced anisotropy energy (MsA) is given as: $E_{MsA} = -K_{MsA} \cos^2 \theta$ [1][18], where $K_{MsA}$ is given as $\frac{3}{2}\lambda\sigma$. So, energy depends on both the sign of $\sigma\lambda$ and $\theta$, where $\theta$ is the angle between magnetization and stress direction.

For fcc-Py, in the RT regime, $\sigma\lambda$ is positive, which means the energy will be minimum when $\theta$ is 0°. This will result in the IP configuration of magnetization in the demagnetized state, if it is the sole contributor. The sign of $\sigma\lambda$ changes in the ST regime as $\lambda$ is negative in this regime, hence, $\theta = 90°$ is the preferred orientation. On the other hand, stress induced due to growth conditions is compressive in nature; hence, the preferred orientation of magnetization will be opposite to that of lattice-mismatch-induced stress anisotropy for both regimes.

| Regime | RT (Py) | ST (Py) | MIP (Co) | ST (Co) |
|---|---|---|---|---|
| Thickness | d = 5-25 nm | d = 50-125 nm | t = 5-90 nm | t = 100-150 nm |
| Magnetostriction constant, $\lambda$ | + | − | + | + |
| Stress, $\sigma_l$ | + | + | | |
| Stress, $\sigma_g$ | − | − | − | − |
| $\sigma\lambda_l$ | + | − | | |
| $\sigma\lambda_g$ | − | + | − | − |
| M (l) | → | ↑ | | |
| M (g) | ↑ | → | ↑ | ↑ |

FIGURE 11S: A symbolic representation of the behaviour of stress anisotropy components in different regimes, indicating the driving mechanism responsible for such magnetization configuration due to MsA. Subscripts, 'g' and 'l', correspond to 'growth-induced' and 'lattice mismatch'.

The case of a-Co is quite simple, as its amorphous nature leads to the absence of lattice-mismatch-induced anisotropy. The growth-induced stress anisotropy always prefers the OOP direction due to positive $\lambda$.

**References:**

1S. Gayen, A., Prasad, G.K., Mallik, S., Bedanta, S., and Perumal, A. (2017) Effects of composition, thickness and temperature on the magnetic properties of amorphous CoFeB thin films. *J. Alloys Compd.*, **694**, 823–832.

2S. Klokholm, E., and Aboaf, J.A. (1981) The saturation magnetostriction of permalloy films. *J. Appl. Phys.*, **52** (3), 2474–2476.